\documentclass[aps,prd,onecolumn,groupedaddress,showpacs,nofootinbib,amssymb]{revtex4}
\usepackage[dvips]{graphicx}
\usepackage{amssymb}
\usepackage{amsmath}
\usepackage{graphicx}
\usepackage{amsfonts}
\usepackage{bm}
\usepackage{latexsym,float}
\usepackage{epsfig}
\usepackage{epstopdf}
\usepackage{amssymb}
\usepackage{verbatim}
\usepackage{multirow}
\usepackage{color}

\begin{document}

\title{Spin connection and cosmological perturbations in $f(T)$ gravity.}

\author{Alexey V. Toporensky$^{1,2}$}
\email{atopor@rambler.ru} \affiliation{$^{1)}$Sternberg Astronomical Institute, Moscow State University, Moscow 119991, Russia\\ $^{2)}$Institute of Physics, Kazan Federal University,
Kremlevskaya street 18, 420008 Kazan, Russia}

\author{Petr V. Tretyakov$^{2,3}$}
\email{tpv@theor.jinr.ru} \affiliation{$^{3)}$Joint Institute for
Nuclear Research, Joliot-Curie 6, 141980 Dubna, Moscow region,
Russia}

\begin{abstract}
In this paper we study small scalar cosmological perturbations in teleparallel gravity in linear order. We discuss problems which appear in standard approach to $f(T)$ gravity, and find that these problems may be solved within covariant formulation of teleparallel gravity, which take into account spin connection. We calculate spin connection which symmetrize equation for perturbation and split diagonal and non-diagonal part of equation of motion. We demonstrate that there is a solution for spin connection, which lead to zero gravitational  
slip,
however, in this case one additional equation appears, so the system may become over-determined.  After that, we show that a more general solution exists,
which is incompatible with zero slip, but allows to write down the equations of motion for cosmological perturbation in a self-consistent way without additional
equations to be satisfied.

\end{abstract}


\maketitle

\section{Introduction}\label{sec:1}

Recently theories of modified teleparallel gravity became an area of intense investigations. Based on
the Einstein idea (proposed by him a decade after he discovered General Relativity)
to use curvature-free connection instead of torsion-free used in General Relativity (GR),
and, so, to use torsion instead of curvature to describe deviation the metric from the Minkowski one \cite{Einstein},
this theory gives rise to a set of possible modifications of GR which can not be constructed in the
classical curvature formalism. If the torsion scalar is inserted in the action of the theory the same way,
as the curvature scalar $R$ enters in the Hilbert action, the theory appears to be equivalent to GR, and,
therefore is usually called as Teleparallel Equivalent of General Relativity (TEGR) \cite{AP} . If the action is modifies
the same way as we get $f(R)$ theory, the resulting $f(T)$ theory is, however, not equivalent to $f(R)$ theory
and represent a new class of modified gravity theories \cite{FF1,BF1,Linder}.

This class has some nice features, for example, equations of motion are still the second order differential
equation, in contrast to 4-th order in $f(R)$. The list of papers on this theory
is already rather long (see, for example \cite{BMNO,BCNO,RHSGR,BOSG,BCLNSG,N1,CCLS,N2,CLSX,KS1,Tret} and references therein). However, $f(T)$ theories have their own problems, the most
serious being the lack of local Lorentz invariance \cite{LSB,SLB}. To understand the meaning of this problem it is
necessary to remember that while GR and its modification require only metrics, TEGR and its modification
use tetrad fields as a fundamental objects. In TEGR the tetrad used can be locally rotated by Lorentz
transformation without modifying the equations of motion. This property is, however, lost in $f(T)$ theories.
This complicates the analysis of equations of motion significantly. Indeed, in the case of local Lorentz
invariance we can, while studying some physical situation, use the simplest tetrad which can be got from
the appropriate metric ansatz. This simple way is absent if the local Lorentz invariance is lost --
the simplest tetrad usually leads to inconsistent system of equations of motion.

There are two ways to solve this problem. One way is to find a tetrad (called as a proper or good tetrad)
which gives us the consistent system  \cite{TB}. The other way allows us to start with an arbitrary tetrad
(which means that we can choose the simplest possible tetrad), and
introduce new independent gravitational variable, so-called spin connection. This allows to restore Lorentz symmetry in the theory and fix the problem of bad and good tetrads \cite{KS2,G2,HJU,G3}.
 Both ways can be implemented rather easily if the
physical system in question has large number of symmetries. For example, in flat Friedmann cosmology a diagonal
cartesian tetrad is a proper tetrad. However, already for a spherically symmetric case the proper tetrad
appears to be a nondiagonal one which make calculation more cumbersome \cite{arg}. On the other hand, spin connection
for a diagonal tetrad has rather simple form \cite{KS2}.

The less number of symmetries has the physical system, the more cumbersome the calculations become.
Highly symmetrical system, like homogeneous and isotropic space-time gives us a general
picture, but obviously, are not enough if we would like to understand the physical system better.
For example, in cosmology small perturbations on Friedmann background play an important role, since
they can be observed and measured, giving a chance to confront theory and experiment.
That is why
for any modified theory of gravity the question of cosmological perturbations is very important. Due to complicated calculations the number of papers in this field is still very small. It is possible to mention \cite{IO} where was produce attempt to investigate cosmological perturbations in $f(T)$ theory of gravity coupled with scalar field and a few woks on pure $f(T)$ theory \cite{CDDS,DDS,WG,G1}.
 The equation for scalar perturbations was first derived  in \cite{CDDS}. Authors of that paper obtained that equation not symmetric and the system is over-determined (number of equations more then number of variables). The only possibility to fix this problem was by requiring $f''(T)\simeq 0$. To our mind, this mean that
the tetrad used in \cite{CDDS} is not a proper tetrad and should be equipped with appropriate non-zero
spin connections to get correct equations of motion.

In our paper we apply the approach based on spin connections to the problem of cosmological perturbations. We calculate corresponding spin connection in implicit form and derive corresponding equations for scalar perturbations.
This paper is organized as follows: in Sec.2 we remind a reader formulation of $f(T)$ gravity, in particular, covariant formulation
which implies non-zero spin connection. In Sec.3 we write down the equations for spin connections which can close the system of equations of motion for scalar cosmological perturbations in $f(T)$ gravity. In Sec. 4 we 
describe an attempt to find solution of equations for spin connections allowing zero gravitational slip, and outline corresponding difficulties. In Sec. 5 we present a non-zero slip solution for the spin connections. In Sec.6 we briefly summarize our main results.

\section{Standard and covariant formulation of teleparallel gravity.}

\subsection{Standard approach to teleparallel gravity.}\label{sec:2}

In standard formulation of teleparallel gravity the main geometrical object is so-called tetrad field
$e^A_{\,\,\,\,\mu}$,
 defined  in tangent space. The space-time metric tensor can be expressed through the tetrad field as 
\begin{equation}
g_{\mu\nu}=\eta_{AB}e^A_{\,\,\,\,\mu}e^B_{\,\,\,\,\nu},
\label{1.2}
\end{equation}
where $\eta_{AB}=\mathrm{diag}(1,-1,-1,-1)$ is Minkowski matrix.
Here and below capital Latin indexes relate to the tangent space and take the values $0,..,3$,  whereas Greek indexes relate to  space-time and also take the values $0,..,3$ and small Latin indexes $i,k,l,m$ relate to  space-time and take the values $1,..,3$.
The torsion tensor is defined as
\begin{equation}
T^{\lambda}_{\,\,\,\,\mu\nu}\equiv \Gamma^{\lambda}_{\,\,\,\,\nu\mu} -\Gamma^{\lambda}_{\,\,\,\,\mu\nu} = e_A^{\,\,\,\,\lambda}(\partial_{\mu}e^A_{\,\,\,\,\nu}-\partial_{\nu}e^A_{\,\,\,\,\mu}),
\label{1.3}
\end{equation}
where $e_A^{\,\,\,\,\mu}$ denotes inverse tetrad, which satisfies $e_A^{\,\,\,\,\mu}e^A_{\,\,\,\,\nu}=\delta^{\mu}_{\nu}$ and $e_A^{\,\,\,\,\mu}e^B_{\,\,\,\,\mu}=\delta^{B}_{A}$. In addition, the following tensors are usually defined: contorsion tensor, which equals to the difference between the Weitzenb\"{o}ck and Levi-Civita connection
\begin{equation}
K^{\mu\nu}_{\,\,\,\,\,\,\,\,\lambda}\equiv -\frac{1}{2}\left ( T^{\mu\nu}_{\,\,\,\,\,\,\,\,\lambda} -T^{\nu\mu}_{\,\,\,\,\,\,\,\,\lambda} -T_{\lambda}^{\,\,\,\,\mu\nu} \right ),
\label{1.4}
\end{equation}
and the following auxiliary tensor
\begin{equation}
S_{\lambda}^{\,\,\,\,\mu\nu}\equiv (K^{\mu\nu}_{\,\,\,\,\,\,\,\,\lambda}+\delta^{\mu}_{\lambda} T^{\alpha\nu}_{\,\,\,\,\,\,\,\,\alpha} -\delta^{\nu}_{\lambda} T^{\alpha\mu}_{\,\,\,\,\,\,\,\,\alpha})
\label{1.5}
\end{equation}
Note that last tensor and torsion tensor are antisymmetric
on second and third indexes, so\footnote{Note here that all Greek indexes rise and down by using metric, whereas Latin indexes by using Minkowski matrix.} $T_{\lambda(\mu\nu)}=S_{\lambda(\mu\nu)}=0$, whereas for contorsion tensor we have $K_{(\mu\nu)\lambda}=0$.

Now torsion scalar can
be defined as
\begin{equation}
T\equiv \frac{1}{2} S_{\lambda}^{\,\,\,\,\mu\nu}T^{\lambda}_{\,\,\,\,\mu\nu}  =\frac{1}{4}T^{\lambda\mu\nu}T_{\lambda\mu\nu}+\frac{1}{2}T^{\lambda\mu\nu}T_{\nu\mu\lambda}- T_{\lambda\mu}^{\,\,\,\,\,\,\,\,\lambda}T^{\nu\mu}_{\,\,\,\,\,\,\,\,\nu},
\label{1.6}
\end{equation}
and the action of teleparallel gravity in the most general form can be written as
\begin{equation}
S=\frac{1}{2\kappa^2}\int d^4x\, e\, f(T),
\label{1.7}
\end{equation}
where $e=det(e^A_{\,\,\,\,\mu})=\sqrt{-g}$ and $\kappa^2$ is the gravitational constant. Variation of action (\ref{1.7}) with respect to tetrad field give us equation of motion in the following form
\begin{equation}
e^{-1}\partial_{\mu}(eS_A^{\,\,\,\,\nu\mu})f'-e^{\,\,\,\,\lambda}_AT^{\rho}_{\,\,\,\,\mu\lambda}S_{\rho}^{\,\,\,\,\mu\nu}f' + S_A^{\,\,\,\,\mu\nu}\partial_{\mu}(T)f''+\frac{1}{2}e^{\,\,\,\,\nu}_Af=\kappa^2 e^{\,\,\,\,\rho}_A T_{\,\,\,\rho}^{m\,\,\,\,\nu},
\label{1.8}
\end{equation}
which is also may be rewritten in equivalent form \cite{Barrow3}
\begin{equation}
\tilde{G}_{\mu\nu}\equiv f'(\stackrel{\!\!\!\!\!\!\circ}{R_{\mu\nu}}-\frac{1}{2}g_{\mu\nu}\stackrel{\circ}{R}) + \frac{1}{2}g_{\mu\nu}[f(T)-f'T] +f''S_{\nu\mu\lambda}\nabla^{\lambda}T=\kappa^2 T^{m}_{\mu\nu},
\label{1.9}
\end{equation}
where $\stackrel{\!\!\!\!\!\!\circ}{G_{\mu\nu}}\equiv \stackrel{\!\!\!\!\!\!\circ}{R_{\mu\nu}}-\frac{1}{2}g_{\mu\nu}\stackrel{\circ}{R}$ should
be calculated by using Levi-Civita connection $\stackrel{\!\!\!\!\!\!\!\!\!\!\!\circ}{\Gamma^{\rho}_{\,\,\,\,\mu\nu}}=\frac{1}{2}g^{\rho\sigma}(\partial_{\mu}g_{\sigma\nu}+\partial_{\nu}g_{\mu\sigma} -\partial_{\sigma}g_{\mu\nu})$.

This formulation of teleparallel gravity has well known problems of lack of local Lorentz invariance.
Due to this property, for example, we need to use only cartesian tetrad in FRW cosmology, and can not use a spherically
symmetric tetrad. The proper tetrad we need to use is usually {\it a priori} unknown. 
In the next subsection we remind a reader
one of the way to solve this problem.

\subsection{Covariant formulation of teleparallel gravity.}\label{sec:3}

The problem of proper tetrads can be eliminated in a modern
approach  which is called as covariant formulation of teleparallel gravity. Within this approach instead of definition of Weitzenb\"{o}ck connection (\ref{1.3}) more general expression is used
\begin{equation}
\Gamma^{\lambda}_{\,\,\,\,\mu\nu}=e_A^{\,\,\,\,\lambda}\left ( \partial_{\nu}e^A_{\,\,\,\,\mu} +\omega^{A}_{\,\,\,\,B\nu} e^B_{\,\,\,\,\mu}  \right),
\label{1.10}
\end{equation}
where $\omega^{A}_{\,\,\,\,B\nu}$ is so called spin connection.

One of the important consequence of this approach is possibility to "restore" Lorenz symmetry in the theory
in the sense that we can use any tetrad connected with the proper tetrad by local Lorenz transformation. The price 
for this is in general non-zero spin connection.  We should mention that a spin connection 
in teleparallel gravity should
 satisfy  the set of important conditions. First of all
 it is antisymmetric with respect to Latin indexes since the connection in teleparallel gravity belongs to the Lorentz algebra:
\begin{equation}
\omega^{AB}_{\,\,\,\,\,\,\,\,\mu}=-\omega^{BA}_{\,\,\,\,\,\,\,\,\mu},
\label{1.11}
\end{equation}
where Latin indexes both must be in upper or lower position. Second,
spin connection in teleparallel gravity are flat in the sense that
\begin{equation}
R^{A}_{\,\,\,\,B\mu\nu}(\omega^A_{\,\,\,\,B\mu})=\partial_{\mu}\omega^{A}_{\,\,\,\,B\nu} -\partial_{\nu}\omega^{A}_{\,\,\,\,B\mu}+\omega^{A}_{\,\,\,\,C\mu}\omega^{C}_{\,\,\,\,B\nu} -\omega^{A}_{\,\,\,\,C\nu}\omega^{C}_{\,\,\,\,B\mu}=0.
\label{1.12}
\end{equation}

All this philosophy assumes that tetrad, which correspond to our metric is not defined uniquely, but only up to a local Lorenz transformation, which transforms the spin connection as well \cite{HJU}
\begin{equation}
e'^A_{\,\,\,\,\mu}=\Lambda^A_{\,\,\,\,B} e^B_{\,\,\,\,\mu},\,\,\,\, \omega'^A_{\,\,\,\,\,\,\,\,B\mu}=\Lambda^A_{\,\,\,\,C}\omega^{C}_{\,\,\,\,F\mu}\Lambda_B^{\,\,\,\,F} +\Lambda^A_{\,\,\,\,C}\partial_{\mu}\Lambda_B^{\,\,\,\,C},
\label{1.13}
\end{equation}
where $\Lambda_A^{\,\,\,\,B}$ is the inverse of the Lorenz transformation matrix $\Lambda^A_{\,\,\,\,B}$. The third condition to be satisfied is the
 metricity condition
\begin{equation}
\partial_{\mu} e^A_{\,\,\,\,\nu}+ \omega^{A}_{\,\,\,\,B\mu} e^B_{\,\,\,\,\nu} -\Gamma^{\lambda}_{\,\,\,\,\nu\mu}e^A_{\,\,\,\,\lambda}=0.
\label{1.13.1}
\end{equation}
We do not consider here theories with non-metricity, like symmetric teleparallel gravity.

All definitions (\ref{1.3})-(\ref{1.6}) are unchanged, for instance we have
\begin{equation}
T^{\lambda}_{\,\,\,\,\mu\nu}\equiv \Gamma^{\lambda}_{\,\,\,\,\nu\mu} -\Gamma^{\lambda}_{\,\,\,\,\mu\nu},\,\,\,\,K^{\lambda}_{\,\,\,\,\mu\nu}\equiv   \Gamma^{\lambda}_{\,\,\,\,\mu\nu} -\stackrel{\!\!\!\!\!\!\!\!\!\!\!\circ}{\Gamma^{\lambda}_{\,\,\,\,\mu\nu}},
\end{equation}
however, all tensors  depend on spin connection implicitly through (\ref{1.10}).

This formulation of theory assumes that spin connection and tetrad are independent gravitational variables. For this reason variation with respect to tetrad field gives us the same  result and equation (\ref{1.9}) is unchanged. However, there is additional equation of motion, which appears due to  variation with respect to spin connection:

\begin{equation}
\partial_{\mu}f_T \left [ \partial_{\nu}(ee_{[A}^{\,\,\,\,\,\mu}e_{B]}^{\,\,\,\,\,\nu}) +2ee_C^{\,\,\,\,[\mu}e_{[A}^{\nu]}\omega^{C}_{\,\,\,\,B]\nu} \right ]=0,
\label{1.14}
\end{equation}
and this equation may be interpreted as equation for spin connection.

We should note here that described above modification of the theory can solve the problem of proper tetrads only if
 for any arbitrary tetrad there exist corresponding spin connection, which allows us to  avoid any problems with equations of motion. Still this statement have not been proved, though it is usually implicitly assumed in the ongoing works
 on covariant teleparallel gravity.
 Nevertheless in the most general case solution of equation (\ref{1.14}) may be a rather difficult problem because actually it is a system of equations with partial derivatives (the term  $\partial_{\mu}T$ contains derivatives from spin connection  implicitly). For this reason the correct form of spin connection was found only for a few simplest cases such as Minkowski and FRW metric in spherical coordinate system, static spherically symmetric metric \cite{KS2}. We exclude from this list a number of cases where spin connection may be set to zero without any pathology in equations, such as FRW in Cartesian reference frame.
 
 So that, the problem to find an appropriate spin connections in covariant version  of the theory and the problem of proper tetrads in the older version seems to be at the same level of difficulty. There are no universal methods applicable to any physical system yet,
 though certain proposals based on underlying symmetries \cite{Manuel} and "switching off" gravity \cite{Petrov} are currently
 under discussions. The goal of the present paper is to show that the appropriate spin connection can be found in a 
 much less symmetric system  (in comparison with systems studied earlier), in almost algorithmic way of calculations
 without any guessing (so, in some sense, by "brute force"), which gives us hope that new systems with less symmetries will
 be added 
 list of known appropriate
 spin connections soon.

It was proved \cite{G3} that equation (\ref{1.14}) coincides  with non-symmetric part of equation of motion in standard formulation of teleparallel gravity, and therefore in the covariant formulation of teleparallel gravity the equation of motion is symmetric automatically. Due to this fact \cite{LSB} the following geometrical identity is satisfied:
\begin{equation}
\stackrel{\!\!\!\!\circ}{\nabla_{\mu}}\tilde{G}^{\mu}_{\nu}=0,
\end{equation}
and therefore the standard (similar to GR) conservation equation is realized $\stackrel{\!\!\!\!\circ}{\nabla_{\mu}}T^m_{\mu\nu}=0$.

\section{First order  cosmological perturbations in teleparallel gravity.}\label{sec:4}

To investigate the problem of cosmological perturbations we start from the perturbed FRW-metric
\begin{equation}
g_{\mu\nu}=diag\big [(1+2\psi), -a^2(1+2\phi), -a^2(1+2\phi), -a^2(1+2\phi)  \big ],
\label{2.1}
\end{equation}
and corresponding tetrad field
\begin{equation}
e^A_{\,\,\,\mu}=diag\big[ 1+\psi, a(1+\phi), a(1+\phi), a(1+\phi)   \big ],
\label{2.2}
\end{equation}
where $a=a(t)$ is scale factor and $\psi=\psi(t,x,y,z)$, $\phi=\phi(t,x,y,z)$ are small perturbations of metric. The inverse
metric and tetrad field matrix may be easily found as
\begin{equation}
g^{\mu\nu}=diag\big [(1-2\psi), -a^{-2}(1-2\phi), -a^{-2}(1-2\phi), -a^{-2}(1-2\phi)  \big ],
\end{equation}
and
\begin{equation}
e_A^{\,\,\,\mu}=diag\big[ 1-\psi, a^{-1}(1-\phi), a^{-1}(1-\phi), a^{-1}(1-\phi)   \big ].
\end{equation}

An attempt to derive equations for cosmological perturbations following this way was done in \cite{CDDS}. Authors
did not take into account spin connection and worked in the logic described in sec.\ref{sec:2}. They found that equations of motion contains non-symmetric non-diagonal part and the only possibility is to put $f''(T)\simeq 0$.  In the present paper
we  correct  this  result  by  taking  into  account  spin connection.  We  should  note  that  construction  of  corresponding
spin  connections  which  satisfy all condition  described  in  sec. II B  is  a  rather  difficult  problem.   Here  we  focus
our attention on a less ambitious task:  to remove non-symmetric part from equations of motions by introducing appropriate spin
connection, calculating explicitly only those components of spin connection which are needed for this goal.

First of all we have the following consideration.  FRW metric in Cartesian reference frame without perturbations
has no any problem with equations of motion and the non-diagonal part is absent (this fact was confirmed by the
authors of \cite{CDDS}) even for zero spin connection. Thus all non-trivial components of spin connection have the
order of perturbations and during the calculations all quadratic and higher order terms may be neglected.  This fact
simplify all calculations seriously, this is the reason why we work in the Cartesian tetrad.

\subsection{Anisymmetry conditions for spin connections.}

We start with antisymmetry condition (\ref{1.11}) for spin connection. It means that in the most general case there are $24$ independent non-trivial components (all $\omega^A_{\,\,\,\,A\mu}=0$).

The independent $\omega^A_{\,\,\,\,B\mu}$ are:

\begin{equation}
\begin{array}{c}
\omega^0_{\,\,\,10},\,\,\omega^0_{\,\,\,11},\,\,\omega^0_{\,\,\,12},\,\,\omega^0_{\,\,\,13},\,\, \omega^0_{\,\,\,20},\,\,\omega^0_{\,\,\,21},\,\,\omega^0_{\,\,\,22},\,\,\omega^0_{\,\,\,23},\,\,
\omega^0_{\,\,\,30},\,\,\omega^0_{\,\,\,31},\,\,\omega^0_{\,\,\,32},\,\,\omega^0_{\,\,\,33},\,\,\\
\\ \omega^1_{\,\,\,20},\,\,\omega^1_{\,\,\,21},\,\,\omega^1_{\,\,\,22},\,\,\omega^1_{\,\,\,23},\,\,
\omega^1_{\,\,\,30},\,\,\omega^1_{\,\,\,31},\,\,\omega^1_{\,\,\,32},\,\,\omega^1_{\,\,\,33},\,\,
\omega^2_{\,\,\,30},\,\,\omega^2_{\,\,\,31},\,\,\omega^2_{\,\,\,32},\,\,\omega^2_{\,\,\,33},\,\,
\end{array}
\end{equation}

and the dependent are

\begin{equation}
\begin{array}{c}
\omega^1_{\,\,\,00}=\omega^0_{\,\,\,10},\,\,\omega^1_{\,\,\,01}=\omega^0_{\,\,\,11},\,\,\omega^1_{\,\,\,02}=\omega^0_{\,\,\,12}, \,\,\omega^1_{\,\,\,03}=\omega^0_{\,\,\,13},\,\, \omega^2_{\,\,\,00}=\omega^0_{\,\,\,20},\,\,\omega^2_{\,\,\,01}=\omega^0_{\,\,\,21},\,\,\\
\\ \omega^2_{\,\,\,02}=\omega^0_{\,\,\,22},\,\,\omega^2_{\,\,\,03}=\omega^0_{\,\,\,23},\,\,
\omega^3_{\,\,\,00}=\omega^0_{\,\,\,30},\,\,\omega^3_{\,\,\,01}=\omega^0_{\,\,\,31},\,\,\omega^3_{\,\,\,02}=\omega^0_{\,\,\,32}, \,\,\omega^3_{\,\,\,03}=\omega^0_{\,\,\,33},\,\,\\
\\ \omega^2_{\,\,\,10}=-\omega^1_{\,\,\,20},\,\,\omega^2_{\,\,\,11}=-\omega^1_{\,\,\,21},\,\,\omega^2_{\,\,\,12}=-\omega^1_{\,\,\,22}, \,\,\omega^2_{\,\,\,13}=-\omega^1_{\,\,\,23},\,\,
\omega^3_{\,\,\,10}=-\omega^1_{\,\,\,30},\,\,\omega^3_{\,\,\,11}=-\omega^1_{\,\,\,31},\,\,\\
\\ \omega^3_{\,\,\,12}=-\omega^1_{\,\,\,32},\,\,\omega^3_{\,\,\,13}=-\omega^1_{\,\,\,33},\,\,
\omega^3_{\,\,\,20}=-\omega^2_{\,\,\,30},\,\,\omega^3_{\,\,\,21}=-\omega^2_{\,\,\,31},\,\,\omega^3_{\,\,\,22}=-\omega^2_{\,\,\,32}, \,\,\omega^3_{\,\,\,23}=-\omega^2_{\,\,\,33},
\end{array}
\end{equation}
all this values depend on all coordinates in the most general case.

Now taking tetrad (\ref{2.2}) and all $24$ independent components of $\omega^A_{\,\,\,\,B\mu}$ we can calculate all values which is needed for equation of motion (\ref{1.9}), keeping only linear with respect to $\psi$, $\phi$ and $\omega^A_{\,\,\,\,B\mu}$ (and its derivatives) terms.

It is need to say here that expressions for $T^{\lambda}_{\,\,\,\,\mu\nu}$ and $S^{\lambda}_{\,\,\,\,\mu\nu}$ take a lot of place even in linearized form, so we do not present them here. We write down here final result for torsion scalar and for $S_{ikl}\partial^l T$ product because they are
important for equations of motion.

Calculation for torsion scalar give us the following result (we keep here only linear terms with respect to perturbations)
\begin{equation}
T=-6H^2+12H^2\psi -12H\dot\phi + 4\frac{H}{a}\big (\omega^0_{\,\,\,11}+\omega^0_{\,\,\,22}+\omega^0_{\,\,\,33}  \big),
\label{2.3}
\end{equation}
and  we  can  see  that  in  spite  of  most  general  form  of  spin  connection  only  three  components  are  present  in  this
expression.

\subsection{Flatness  conditions for spin connections.}

Now we consider flatness conditions of spin connections. We can neglect two last terms in (\ref{1.12}), because they are
second order in perturbations.  Thus the conditions can be presented in the following form
\begin{equation}
\partial_{\mu}\omega^A_{\,\,\,B\nu}=\partial_{\nu}\omega^A_{\,\,\,B\mu}.
\end{equation}
Detailed analysis shows us that solutions of this system of differential equations can be written as:

\begin{eqnarray}
&&\omega^0_{\,\,\,10}= a_{1t},\,\,\,\omega^0_{\,\,\,11}= a_{1x},\,\,\,\omega^0_{\,\,\,12}= a_{1y},\,\,\,\omega^0_{\,\,\,13}= a_{1z},
\label{2.4.18}\\
&&\omega^0_{\,\,\,20}= a_{2t},\,\,\,\omega^0_{\,\,\,21}= a_{2x},\,\,\,\omega^0_{\,\,\,22}= a_{2y},\,\,\,\omega^0_{\,\,\,23}= a_{2z},
\label{2.4.19}\\
&&\omega^0_{\,\,\,30}= a_{3t},\,\,\,\omega^0_{\,\,\,31}= a_{3x},\,\,\,\omega^0_{\,\,\,32}= a_{3y},\,\,\,\omega^0_{\,\,\,33}= a_{3z},
\label{2.4.20}\\
&&\omega^1_{\,\,\,20}=  b_{1t},\,\,\,\omega^1_{\,\,\,21}= b_{1x},\,\,\,\omega^1_{\,\,\,22}= b_{1y},\,\,\,\omega^1_{\,\,\,23}= b_{1z},
\label{2.4.21}\\
&&\omega^1_{\,\,\,30}= b_{2t},\,\,\,\omega^1_{\,\,\,31}= b_{2x},\,\,\,\omega^1_{\,\,\,32}= b_{2y},\,\,\,\omega^1_{\,\,\,33}= b_{2z},
\label{2.4.22}\\
&&\omega^2_{\,\,\,30}=  c_{t},\,\,\,\,\,\omega^2_{\,\,\,31}= c_{x},\,\,\,\,\,\omega^2_{\,\,\,32}= c_{y},\,\,\,\,\,\omega^2_{\,\,\,33}= c_{z},
\label{2.4.23}
\end{eqnarray}
where $a_i$, $b_i$ and $c$ are some functions of the coordinates $(t,x,y,z)$ and index $\mu=t,x,y,z$ denotes differentiation with respect to coordinate.

\subsection{Non-symmetric part of equations of motion.}

Consider again the equation of motion (\ref{1.9}). Any non-symmetric term can arise from the only source $M_{\nu\mu}\equiv S_{\nu\mu\lambda}\nabla^{\lambda}T$. Calculating its components up to the first order in perturbations we have:

\begin{eqnarray}
M_{00}=&&0,
\label{2.4.1}\\
M_{01}=&&12\frac{\dot a}{a^3}\big (a\ddot a- \dot a^2\big )\big ( \omega^1_{\,\,\,22} + \omega^1_{\,\,\,33} +2\phi_x \big ) ,
\label{2.4.2}\\
M_{02}=&& 12\frac{\dot a}{a^3}\big ( a\ddot a- \dot a^2\big )\big ( \omega^2_{\,\,\,33} - \omega^1_{\,\,\,21} +2\phi_y \big ),
\label{2.4.3}\\
M_{03}=&& 12\frac{\dot a}{a^3}\big ( a\ddot a- \dot a^2\big )\big ( -\omega^1_{\,\,\,31} - \omega^2_{\,\,\,32} +2\phi_z \big ),
\label{2.4.4}\\
M_{10}=&& -8\frac{\dot a^2}{a^3} \frac{\partial}{\partial x} \Big ( \omega^0_{\,\,\,11}+\omega^0_{\,\,\,22}+\omega^0_{\,\,\,33} - 3a\phi_t+3\dot a\psi \Big),
\label{2.4.5}\\
M_{11}=&& M_0+ M_* + 12\frac{H}{a} (a\ddot a- \dot a^2)\omega^0_{\,\,\,11}  ,
\label{2.4.6}\\
M_{12}=&& 12\frac{\dot a}{a^2}\big ( a\ddot a- \dot a^2\big ) \omega^0_{\,\,\,21},
\label{2.4.7}\\
M_{13}=&&  12\frac{\dot a}{a^2}\big ( a\ddot a- \dot a^2\big ) \omega^0_{\,\,\,31},
\label{2.4.8}\\
M_{20}=&& -8\frac{\dot a^2}{a^3} \frac{\partial}{\partial y} \Big ( \omega^0_{\,\,\,11}+\omega^0_{\,\,\,22}+\omega^0_{\,\,\,33} - 3a\phi_t+3\dot a\psi \Big),
\label{2.4.9}\\
M_{21}=&&  12\frac{\dot a}{a^2}\big ( a\ddot a- \dot a^2\big ) \omega^0_{\,\,\,12},
\label{2.4.10}\\
M_{22}=&& M_0+ M_*  + 12\frac{H}{a} (a\ddot a- \dot a^2)\omega^0_{\,\,\,22} ,
\label{2.4.11}\\
M_{23}=&&  12\frac{\dot a}{a^2}\big ( a\ddot a- \dot a^2\big ) \omega^0_{\,\,\,32},
\label{2.4.12}\\
M_{30}=&& -8\frac{\dot a^2}{a^3} \frac{\partial}{\partial z} \Big ( \omega^0_{\,\,\,11}+\omega^0_{\,\,\,22}+\omega^0_{\,\,\,33} - 3a\phi_t+3\dot a\psi \Big),
\label{2.4.13}\\
M_{31}=&&  12\frac{\dot a}{a^2}\big ( a\ddot a- \dot a^2\big ) \omega^0_{\,\,\,13},
\label{2.4.14}\\
M_{32}=&&  12\frac{\dot a}{a^2}\big ( a\ddot a- \dot a^2\big ) \omega^0_{\,\,\,23},
\label{2.4.15}\\
M_{33}=&& M_0+M_*+ 12\frac{H}{a} (a\ddot a- \dot a^2)\omega^0_{\,\,\,33}  ,
\label{2.4.16}
\end{eqnarray}
with


\begin{eqnarray}
M_0=&&-24\Big[ H^2(\dot a^2-a\ddot a)\Big] - 24\Big[ H^2a\dot a\dot\psi -4H^2(\dot a^2-a\ddot a)\psi +2H^2(\dot a^2-a\ddot a)\phi +2H(\dot a^2-a\ddot a)\dot\phi -\dot a^2\ddot\phi \Big]\label{2.4.17.1}\\
M_*=&&4\frac{H}{a}\Big[ (7\dot a^2-5a\ddot a)(\omega^0_{\,\,\,11}+\omega^0_{\,\,\,22}+\omega^0_{\,\,\,33}) -2a\dot a  \frac{\partial}{\partial t}(\omega^0_{\,\,\,11}+\omega^0_{\,\,\,22}+\omega^0_{\,\,\,33})   \Big].
\label{2.4.17.2}
\end{eqnarray}

\section{Zero slip solution}

A solution of described above systems (\ref{2.4.1})-(\ref{2.4.16}) and (\ref{2.4.18})-(\ref{2.4.23}) allows us to solve our task of construction symmetric equations with flat spin connection.  In the present section we will not discuss the problem of uniqueness of the solution
(which is a more complicated task), but present here a particular solution.

Our first attempt is to find a solution as simple as possible. First of all note, that in the most general case diagonal equations  (\ref{2.4.6}), (\ref{2.4.11}) and (\ref{2.4.16}) are different.  It is reasonable  to make them the same  by setting $\omega^0_{\,\,\,11}=\omega^0_{\,\,\,22}=\omega^0_{\,\,\,33}\neq 0$. Next, the conditions involving $\omega$ with upper index $^0$ decouple, giving
$a_{1x}=a_{2y}=a_{3z}$, $a_{1y}=a_{2x}$, $a_{2z}=a_{3y}$, $a_{1z}=a_{3x}$. To maximally simplify our task we put $a_1=a_2=a_3=0$. In this case we will have only three equations from symmetry conditions: $M_{01}=M_{10}$, $M_{02}=M_{20}$ and $M_{03}=M_{30}$, which contain six variables: $\omega^1_{\,\,\,22}$, $\omega^1_{\,\,\,33}$, $\omega^1_{\,\,\,21}$, $\omega^1_{\,\,\,31}$, $\omega^2_{\,\,\,32}$ and $\omega^2_{\,\,\,33}$. These  variables may be parametrized according to (\ref{2.4.21})-(\ref{2.4.23}) by  three independent functions $b_1$, $b_2$ and $c$. Thus finally we have the following  system for unknown parameters (note also that  these equations may be derived from (\ref{1.14}) by direct calculations):
 
\begin{eqnarray}
\big( a\ddot a-\dot a^2 \big)\big( b_{1y}+b_{2z} \big)=&& 2a\dot a\phi_{tx} +2\big(\dot a^2-a\ddot a\big)\phi_x-2\dot a^2\psi_x,
\label{2.4.24}\\
\big( a\ddot a-\dot a^2 \big)\big( c_{z}-b_{1x} \big)=&& 2a\dot a\phi_{ty} +2\big(\dot a^2-a\ddot a\big)\phi_y-2\dot a^2\psi_y,
\label{2.4.25}\\
-\big( a\ddot a-\dot a^2 \big)\big( b_{2x}+c_{y} \big)=&& 2a\dot a\phi_{tz} +2\big(\dot a^2-a\ddot a\big)\phi_z-2\dot a^2\psi_z.
\label{2.4.26}
\end{eqnarray}

Finally we conclude that solution $b_1=b_1(t,x,y,z)$, $b_2=b_2(t,x,y,z)$, $c=c(t,x,y,z)$ of the system (\ref{2.4.24})-(\ref{2.4.26}) parametrize spin connection totally and produce zero Riemann tensor and symmetric equations of motion under construction up to the first order of perturbations.  Also it is interesting to note, that found spin connection in the minimal form does not affect the diagonal part of equations of motion.

Now in this case we have for $M_{\mu\nu}$

\begin{eqnarray}
M_{00}&&=0,
\label{2.4.27}\\
M_{ii}&&= M_0 ,
\label{2.4.28}\\
M_{0i}&&= M_{i0}=  24H^2 \partial_i\big ( \dot\phi -H\psi \big ),
\label{2.4.29}\\
M_{ij}&& \underset{i\neq j}{=} 0.
\label{2.4.30}
\end{eqnarray}
We can see that $i\neq j$ components of equation (\ref{1.9}) take the form
\begin{equation}
f'\stackrel{\!\!\!\!\!\!\circ}{R_{ij}}=-f'\partial_i\partial_j (\phi+\psi)=\kappa^2\pi_{ij},
\label{2.4.31}
\end{equation}
where $\pi_{ij}$ is anisotropic stress tensor. We will discuss the case of zero anisotropic stress only (which is natural for perfect fluid case), thus we have condition
\begin{equation}
\psi=-\phi,
\label{2.4.32}
\end{equation}
which is the same as in classical GR. 
Note that in many modified gravity theories two scalar potentials $\phi$ and $\phi$ are in general different. The 
difference is called a gravitational slip which is usually parametrized by the value $\eta = |\phi/ \psi|$,
so zero slip corresponds to $\eta=1$ \cite{Sawicki}. Non-zero gravitational slip also appears beyond first order
in perturbations even in the classical GR \cite{Ballesteros}.

Using this solution we can consider the corresponding form of equations of motion.
First of all note that we write equations for perturbations by using (\ref{1.9}) with one upper and one lower indexes, which is more conventional  for this task.

\begin{equation}
\tilde{G}^{\mu}_{\nu}\equiv f'(\stackrel{\!\!\!\circ}{R^{\mu}_{\nu}}-\frac{1}{2}\delta^{\mu}_{\nu}\stackrel{\circ}{R}) + \frac{1}{2}\delta^{\mu}_{\nu}[f(T)-f'T] +f''M^{\mu}_{\nu}=\kappa^2 T^{\mu}_{\nu}.
\label{2.5.1}
\end{equation}

Energy-momentum tensor has the following form

\begin{equation}
T^{\mu}_{\nu}= (\rho+p)u^{\mu}u_{\nu}-\delta^{\mu}_{\nu}p,
\label{2.5.2}
\end{equation}
and
\begin{eqnarray}
T^0_0&&= \delta\rho, \label{2.5.3}\\
\nonumber\\
T^0_i&&= -(p+\rho)\partial_i\delta u, \label{2.5.4}\\
\nonumber\\
T^i_j&&= -\delta^i_j\delta p, \label{2.5.5}
\end{eqnarray}

\vspace{1cm}

Non-trivial components of Einstein tensor  $
\stackrel{\!\!\!\circ}{G^{\mu}_{\nu}}\equiv\stackrel{\!\!\!\circ}{R^{\mu}_{\nu}}-\frac{1}{2}\delta^{\mu}_{\nu}\stackrel{\circ}{R}
$ read

\begin{equation}
\stackrel{\!\!\!\circ}{G^{0}_{0}}=3H^2+ 2 \left ( 3H^2\phi +3H\dot\phi -\frac{1}{a^2}\triangle \phi  \right ),
\label{2.5.6}
\end{equation}

\begin{equation}
\stackrel{\!\!\!\circ}{G^{0}_{i}}= -2\partial_i\left( \dot\phi +H\phi  \right ),
\label{2.5.7}
\end{equation}

\begin{equation}
\stackrel{\!\!\!\circ}{G^{i}_{i}}=2\dot H +3H^2 +2\left ( \ddot\phi +4H\dot\phi +3H^2\phi + 2\dot H\phi   \right ),
\label{2.5.8}
\end{equation}


Non-trivial components of $M^{\mu}_{\nu}$ tensor are

\begin{equation}
M^0_{i} = 24H^2 \partial_i\big ( \dot\phi + H\phi \big ),
\label{2.5.9}
\end{equation}

\begin{equation}
M^i_i= -24H^2\dot H  - 24 H \big[ H\ddot\phi + H^2\dot\phi +2\dot H \dot\phi +4\dot HH\phi \big].
\label{2.5.10}
\end{equation}

Finally decomposing derivatives of function $f$ in Taylor series near the point $T=T_0\equiv -6H^2$ we have

\begin{equation}
f(T)=f(T_0)+f'(T_0)\delta T,
\end{equation}

\begin{equation}
f'(T)=f'(T_0)+f''(T_0)\delta T,
\end{equation}

\begin{equation}
f''(T)=f''(T_0)+f'''(T_0)\delta T,
\end{equation}

where we imply $f'(T_0)=f'(T)|_{T=T_0}\equiv f'_0$ and so on and according to (\ref{2.3}) $\delta T= -12H^2\phi -12H\dot\phi$.

Collecting all terms we have

\begin{eqnarray}
\tilde{G}^0_0=&&    2f'_0 \left ( 3H^2\phi +3H\dot\phi -\frac{1}{a^2}\triangle \phi  \right ) - 72H^3f''_0  \big ( \dot\phi +H\phi    \big )= \kappa^2\delta\rho, \label{2.5.11}\\
\nonumber\\
\tilde{G}^0_{i}=&& -2(f'_0 - 12 f''_0H^2) \partial_i\big ( \dot\phi + H\phi \big )= -\kappa^2(p+\rho)\partial_i\delta u, \label{2.5.12}\\
\nonumber\\
\tilde{G}^i_i =&&  2 f'_0 \left ( \ddot\phi +4H\dot\phi +3H^2\phi + 2\dot H\phi   \right ) +288H^3\dot Hf'''_0 (\dot\phi +H\phi)\label{2.5.13}\\&&  -24Hf''_0 \left( H\ddot\phi +3\dot H\dot\phi  +4H^2\dot\phi +5\dot HH\phi +3H^3\phi  \right )= -\kappa^2\delta p, \nonumber
\end{eqnarray}
calculation of the divergence of this tensor give us $\stackrel{\!\!\!\!\circ}{\nabla_{\mu}}\tilde{G}^{\mu}_{\nu}=0$ as it should to be (note that for this calculation we need to take into account also the condition (\ref{3.11.1}), see below).

\section{Solution with non zero slip} \label{subsec5}

We should stress that  despite we does not need the full solution of the system for the spin connection, we need to know that this
system has some nontrivial solution.  Solution for spin connection described above may be absent for the following
reason.  System (\ref{2.4.24})-(\ref{2.4.26})  can be rewritten in the form

\begin{equation}
\mathbf{rot} \overrightarrow{B}=\mathbf{grad}\Phi,
\label{grad}
\end{equation}
where $\overrightarrow{B}$ is the 3d vector with the components $\overrightarrow{B}=(a\ddot a-\dot a^2)(c,\,\,-b_2,\,\,b_1)$ and $\Phi =2a\dot a\phi_{t} +2\big(\dot a^2-a\ddot a\big)\phi-2\dot a^2\psi$. After taking the divergence from left and right sides of the previous equation we find the following relation for perturbations only (here we imply $\psi=-\phi$)
\begin{equation}
\triangle(H\dot\phi +H^2\phi-\dot H\phi)=0.
\label{3.11.1}
\end{equation}

If this relation is not satisfied automatically the system becomes over-determined.
This situation is similar to described in \cite{Jarv} where an attempt to construct spin connections for rotating black hole
solution leads to an additional equation. We think that this may appear due to oversimplification of the ansatz chosen, and we should use more general
form of spin connection to avoid this problem. So that we take
 $a_1\neq 0$, $a_2\neq 0$, $a_3\neq 0$. In this case we need to imply the following symmetry
conditions\footnote{We are forced to break condition $\omega^0_{\,\,\,11}=\omega^0_{\,\,\,22}=\omega^0_{\,\,\,33}\neq 0$ (or $a_{1x}=a_{2y}=a_{3z}$) in this more general case.} $\omega^0_{\,\,\,21}=\omega^0_{\,\,\,12}$, $\omega^0_{\,\,\,31}=\omega^0_{\,\,\,13}$, $\omega^0_{\,\,\,32}=\omega^0_{\,\,\,23}$ (or $a_{1y}=a_{2x}$, $a_{2z}=a_{3y}$, $a_{1z}=a_{3x}$). Now instead of system (\ref{2.4.24})-(\ref{2.4.26}) we have

\begin{eqnarray}
\big(a\ddot a- \dot a^2 \big)\big( b_{1y}+b_{2z} \big)=&& 2a\dot a\dot\phi_{x} +2\big(\dot a^2-a\ddot a\big)\phi_x-2\dot a^2\psi_x-\frac{2}{3}\dot a\omega^0_x,
\label{3.12}\\
\big( a\ddot a- \dot a^2 \big)\big( c_{z}-b_{1x} \big)=&& 2a\dot a\dot\phi_{y} +2\big(\dot a^2-a\ddot a\big)\phi_y-2\dot a^2\psi_y-\frac{2}{3}\dot a\omega^0_y,
\label{3.13}\\
-\big( a\ddot a- \dot a^2 \big)\big( b_{2x}+c_{y} \big)=&& 2a\dot a\dot\phi_{z} +2\big(\dot a^2-a\ddot a\big)\phi_z-2\dot a^2\psi_z-\frac{2}{3}\dot a\omega^0_z,
\label{3.14}
\end{eqnarray}
where we denoted
\begin{equation}
\omega^0\equiv \omega^0_{\,\,\,11}+\omega^0_{\,\,\,22}+\omega^0_{\,\,\,33}=a_{1x}+a_{2y}+a_{3z}.
\label{3.15}
\end{equation}
Now the equation (\ref{grad}) is still valid with the new form of $\Phi =2a\dot a\phi_{t} +2\big(\dot a^2-a\ddot a\big)\phi-2\dot a^2\psi - \frac{2}{3}\dot a \omega^0$, and 
after similar mathematical manipulations we find instead of (\ref{3.11.1})
\begin{equation}
\triangle(H\dot\phi -H^2\psi-\dot H\phi-\frac{1}{3}\frac{H}{a}\omega^0)=0.
\label{3.16}
\end{equation}
We can see that now this is not a relation for perturbations only, but some equation connecting components of spin
connection and perturbation potentials which allows us to determine the relation between $\phi$ and $\psi$ (slip) if the spin
connection is known.

Now in this case instead of (\ref{2.4.27})-(\ref{2.4.30}) we have for $M_{\mu\nu}$

\begin{eqnarray}
M_{00}&&=0,
\label{3.17}\\
M_{ii}&&= M_0 -8H\dot a\dot\omega^0 +28 H^2\dot a\omega^0 - 20H\ddot a\omega^0 +12\frac{H}{a}(a\ddot a-\dot a^2)\omega^0_{\,\,\,ii},
\label{3.18}\\
M_{0i}&&= M_{i0}=  24H^2 \partial_i\big ( \dot\phi -H\psi -\frac{1}{3a}\omega^0  \big ),
\label{3.19}\\
M_{ij}&& \underset{i\neq j}{=} 12H(\ddot a- H\dot a) \omega^0_{\,\,\,ij}.
\label{3.20}
\end{eqnarray}
Note here that in our case $M^0_i=M_{0i}$ and $M^i_i=g^{ii}M_{ii}=a^{-2}(2\phi-1)M_{ii}$.
Now we can see that $\{ij\}$-component of equation takes the form
\begin{equation}
f'_0\stackrel{\!\!\!\!\!\!\circ}{R_{ij}}+f''_0M_{ij} \underset{i\neq j}{=} -f'_0\partial_i\partial_j (\phi+\psi)+ 12f''_0H(\ddot a- H\dot a) \omega^0_{\,\,\,ij}=\kappa^2\pi_{ij},
\label{3.21}
\end{equation}
and if we imply zero anisotropic stress, this equation allow us to determine $\omega^0_{\,\,\,ij}$-components($i\neq j$) of spin connection. Note that non-zero $\omega^0_{\,\,\,ij}$
leads to non-zero slip. This confirms the similar result of \cite{G1} and indicates that
non-zero slip can be an intrinsic feature of perturbations in $f(T)$ gravity.

We have now
\begin{equation}
\stackrel{\!\!\!\circ}{G^{0}_{0}}=3H^2+ 2 \left ( 3H\dot\phi -3H^2\psi -\frac{1}{a^2}\triangle \phi  \right ),
\label{3.22}
\end{equation}

\begin{equation}
\stackrel{\!\!\!\circ}{G^{0}_{i}}= -2\partial_i\left( \dot\phi - H\psi  \right ),
\label{3.23}
\end{equation}

\begin{equation}
\stackrel{\!\!\!\circ}{G^{i}_{i}}=2\dot H +3H^2 +2\left ( \ddot\phi +3H\dot\phi -H\dot\psi - 3H^2\psi - 2\dot H\psi   \right ) -\frac{1}{a^2}\left [\triangle(\phi+\psi)-\partial_i\partial_i(\phi+\psi) \right],
\label{3.24}
\end{equation}
and collecting all terms we get
\begin{eqnarray}
\tilde{G}^0_0=&&    2f'_0 \left (3H\dot\phi- 3H^2\psi  -\frac{1}{a^2}\triangle \phi  \right ) - 24H^3f''_0  \left ( 3\dot\phi -3H\psi -\frac{1}{a}\omega^0    \right )= \kappa^2\delta\rho, \label{3.25}\\
\nonumber\\
\tilde{G}^0_{i}=&& -2\left(f'_0 - 12 f''_0H^2\right) \partial_i\left ( \dot\phi - H\psi \right )- 8\frac{H^2}{a}f''_0\partial_i\omega^0= -\kappa^2(p+\rho)\partial_i\delta u, \label{3.26}\\
\nonumber\\
\tilde{G}^i_i =&&  2 f'_0 \left (  \ddot\phi +3H\dot\phi -H\dot\psi - 3H^2\psi - 2\dot H\psi   -\frac{1}{2a^2}\left [\triangle(\phi+\psi)-\partial_i\partial_i(\phi+\psi) \right]\right) \label{3.27}\\&&+288H^3\dot Hf'''_0 (\dot\phi -H\psi -\frac{1}{3a}\omega^0)  -24Hf''_0 \left( H\ddot\phi +3\dot H\dot\phi  +3H^2\dot\phi -H^2\dot\psi -5\dot HH\psi -3H^3\psi  \right )\nonumber\\&& +4\frac{H}{a}f''_0\left( 2H\dot\omega^0 +7\dot H\omega^0 +4H^2\omega^0 \right )-12\frac{H}{a}\dot Hf''_0\omega^0_{\,\,\,ii}= -\kappa^2\delta p. \nonumber
\end{eqnarray}

Now we turn back to equation (\ref{3.21}). We discuss the case of zero anisotropic stress only.  This immediately tell us
that $\omega^0_{\,\,\,ij}=\omega^0_{\,\,\,ji}$ and there are only three independent equations instead of six. They can be written  explicitly through functions $a_i$ (\ref{2.4.18})-(\ref{2.4.20}), this representation allows us later to get also symmetric
components $\omega^0_{\,\,\,ii}$:
\begin{eqnarray}
f'_0\partial_1\partial_2 (\phi+\psi)&&=12f''_0H(\ddot a- H\dot a) a_{1y},
\label{3.28}\\
f'_0\partial_2\partial_3 (\phi+\psi)&&= 12f''_0H(\ddot a- H\dot a) a_{2z} ,
\label{3.29}\\
f'_0\partial_3\partial_1 (\phi+\psi)&&= 12f''_0H(\ddot a- H\dot a) a_{3x},
\label{3.30}
\end{eqnarray}
this may be integrated and rewritten as
\begin{equation}
f'_0\partial_i (\phi+\psi) = 12f''_0H(\ddot a- H\dot a) a_{i}.
\label{3.31}
\end{equation}
We remind a reader that all components of spin connection was parametrized by six functions $a_1$, $a_2$, $a_3$, $b_1$, $b_2$ and $c$ (\ref{2.4.18})-(\ref{2.4.23}). Three of them $a_i$ can be determined using (\ref{3.31}) and exactly this components are needed to close the system of equations. Corresponding components of spin connection can be found after substituting the values of $a_i$ from
(\ref{3.31}) into equations (\ref{2.4.18})-(\ref{2.4.20})
\begin{equation}
\omega^0_{\,\,\,i\mu}=\partial_i\partial_{\mu}\left [\frac{f'_0}{12H(\ddot a-H\dot a)f''_0}(\phi+\psi)\right ],
\label{3.31.1}
\end{equation}
where lower indices can now be equal. This gives us  the value
of  $\omega^0$:
\begin{equation}
\omega^0=\frac{f'_0}{12f''_0H\dot H} \frac{1}{a} \triangle(\phi+\psi).
\label{3.32}
\end{equation}
Substituting  $\omega^0_{\,\,\,ii}$ into Eq.(81) we see that all three  $\tilde{G}^i_i$ become equal to each other as it should be
since r.h.s. of these equations are equal to each other also.

Now we can write down the simplest solution of equation (\ref{3.16})
\begin{equation}
\omega^0= 3a\dot\phi -3aH\psi-3a\frac{\dot H}{H}\phi,
\label{3.32.2}
\end{equation}
it corresponds to   $\mathbf{rot} \overrightarrow{B}=0$, introduced in the beginning of the
present subsection. This indicates that the vector $\overrightarrow{B}$ is a gradient of some function $\alpha$,
and the solution for remaining components of spin connection is
\begin{equation}
c=\partial_x\alpha,\,\,\,b_2=-\partial_y\alpha,\,\,\,b_1=\partial_z\alpha,
\end{equation}
we need not explicit forms of these components for closing the system of equations for perturbations. The function $\alpha$
can be zero, or some other function of the order of perturbations. In particular, if we choose 
$
\alpha=A(\phi+\psi),
$
where 
$
A=\frac{f'_0}{12H(\ddot a-H\dot a)f''_0}
$ is the function of time only,
 we have $
c=A\partial_x(\phi+\psi),\,\,\,b_2=-A\partial_y(\phi+\psi), \,\,\,b_1=A\partial_z(\phi+\psi),$
and according to  (\ref{2.4.18})-(\ref{2.4.23}) and (\ref{3.31.1}) 
we have a very symmetric particular solution for spin connection
\begin{equation}
\omega^1_{\,\,\,2\mu}=\omega^0_{\,\,\,3\mu},\,\,\,\omega^1_{\,\,\,3\mu}=-\omega^0_{\,\,\,2\mu},\,\,\,\omega^2_{\,\,\,3\mu}=\omega^0_{\,\,\,1\mu}.
\end{equation}

Substituting $\omega^0_{\,\,\,ii}$ into the system of equations for perturbations we find finally:
\begin{eqnarray}
\tilde{G}^0_0=&&    2f'_0 \left (3H\dot\phi- 3H^2\psi  -\frac{1}{a^2}\triangle \phi  \right ) - 72H^2\dot H f''_0  \phi= \kappa^2\delta\rho, \label{3.33}\\
\nonumber\\
\tilde{G}^0_{i}=&& -2f'_0\partial_i\left ( \dot\phi - H\psi \right ) + 24H\dot H f''_0\partial_i\phi= -\kappa^2(p+\rho)\partial_i\delta u, \label{3.34}\\
\nonumber\\
\tilde{G}^i_i =&&  2 f'_0 \left (  \ddot\phi +3H\dot\phi -H\dot\psi - 3H^2\psi - 2\dot H\psi   -\frac{1}{2a^2} \triangle(\phi+\psi) \right) \label{3.35}\\&&+288H^2\dot H^2 f'''_0 \phi +12f''_0 \left( -H\dot H\dot\phi  -2H\ddot H\phi -5\dot H^2\phi -6\dot HH^2\phi +\dot HH^2\psi  \right )= -\kappa^2\delta p. \nonumber
\end{eqnarray}
In this system we have four independent variables: $\delta\rho$, $\delta u$, $\phi$ and $\psi$, so we need  one additional equation to close the system. It can be derive by combining (\ref{3.32}) and (\ref{3.32.2}):

\begin{equation}
 36f''_0(H\dot H\dot\phi -H^2\dot H\psi-\dot H^2\phi)=\frac{f'_0}{a^2}\triangle(\phi+\psi).
\label{3.36}
\end{equation}

Thus, we have got a closed system for cosmological perturbation. Straightforward, though rather tedious calculations directly
indicate that $
\stackrel{\!\!\!\!\circ}{\nabla_{\mu}}\tilde{G}^{\mu}_{\nu}=0$ as it should be.

\section{Conclusions.} \label{sec:5}

In the present paper we consider spin connections needed for closing the system of equations for scalar linear
cosmological perturbations in $f(T)$ gravity. It is known that neglecting spin connections lead to the requirement
$f''(T)=0$ which indicated that we can not go beyond GR. Non-zero spin connections can in general improve this situation.
Our first attempt consider the simplest possible spin connections, where all components which can be set to zero are
actually set to zero. We remind a reader that from $64$ components of spin connection $16$ are equal to zero due to antisymmetry,
so in general we have $48$ non-zero components, from which only $24$ are independent.  In the case of minimal solution we have $24$ non-trivial components,
$12$ independent:
\begin{equation}
\begin{array}{c}
\omega^1_{\,\,\,20},\,\,\omega^1_{\,\,\,21},\,\,\omega^1_{\,\,\,22},\,\,\omega^1_{\,\,\,23},\,\,
\omega^1_{\,\,\,30},\,\,\omega^1_{\,\,\,31},\,\,\omega^1_{\,\,\,32},\,\,\omega^1_{\,\,\,33},\,\,
\omega^2_{\,\,\,30},\,\,\omega^2_{\,\,\,31},\,\,\omega^2_{\,\,\,32},\,\,\omega^2_{\,\,\,33},\,\,
\end{array}
\end{equation}

and $12$ dependent:

\begin{equation}
\begin{array}{c}
 \omega^2_{\,\,\,10}=-\omega^1_{\,\,\,20},\,\,\omega^2_{\,\,\,11}=-\omega^1_{\,\,\,21},\,\,\omega^2_{\,\,\,12}=-\omega^1_{\,\,\,22}, \,\,\omega^2_{\,\,\,13}=-\omega^1_{\,\,\,23},\,\,
\omega^3_{\,\,\,10}=-\omega^1_{\,\,\,30},\,\,\omega^3_{\,\,\,11}=-\omega^1_{\,\,\,31},\,\,\\
\\ \omega^3_{\,\,\,12}=-\omega^1_{\,\,\,32},\,\,\omega^3_{\,\,\,13}=-\omega^1_{\,\,\,33},\,\,
\omega^3_{\,\,\,20}=-\omega^2_{\,\,\,30},\,\,\omega^3_{\,\,\,21}=-\omega^2_{\,\,\,31},\,\,\omega^3_{\,\,\,22}=-\omega^2_{\,\,\,32}, \,\,\omega^3_{\,\,\,23}=-\omega^2_{\,\,\,33},
\end{array}
\end{equation}

As a result, we get closed system for the cosmological perturbation which implies zero slip, as in GR.
However, an additional equation for the cosmological perturbation appears, indicating that the system in question becomes
in general over-determined. This forces us to consider more general form of spin connections where all 48 components assumed to be non-zero. We have found
a solution to spin connections equations, which closes the system for the cosmological perturbations, and does not provide any
additional requirements. The price for this is a non-zero slip, arising due to additional non-zero components of spin
connections. It is worth to note that if we choose the function $\alpha$ introduced in Eq.(100)  to be zero,
we get a particular case of non-zero slip case with only 24 non-zero spin connection components,
the same as for the former zero slip case. What is also interesting, all non-zero spin connection
component for zero slip case (listed above in this Section) vanish for this non-zero slip particular case,
and, vice versa, non-zero components in this case are those which are set to zero in the zero slip case.

Our approach is complimentary to the approach of \cite{G1} where the close system for cosmological perturbation in $f(T)$ gravity
have been constructed for a general tetrad with zero spin connections. We leave a detailed comparison of results of these two
approach for a future work. Here we only note that the approach of \cite{G1} also leads to non-zero slip, which may indicate
that these feature is an important peculiarity of $f(T)$ cosmology which can be, in principle, tested experimentally.

\begin{acknowledgments}

This work was supported by the Russian Science Foundation (RSF)
grant 16-12-10401
and partially by the Russian Government Program of Competitive Growth of Kazan Federal University.
Authors are grateful to Alexey Golovnev, Laur J\"arv and Martin Kr\^s\^s\'ak for discussions.

\end{acknowledgments}

\end{document}